\documentstyle[aps,twocolumn]{revtex}

\begin{document}

\input epsf
\newcommand{\infig}[2]{\begin{center}\mbox{ \epsfxsize #1
                       \epsfbox{#2}}\end{center}}

\newcommand{\be}{\begin{equation}}
\newcommand{\ee}{\end{equation}}
\newcommand{\bea}{\begin{eqnarray}}
\newcommand{\eea}{\end{eqnarray}}
\newcommand{\wee}[2]{\mbox{$\frac{#1}{#2}$}}   
\newcommand{\unit}[1]{\,\mbox{#1}}
\newcommand{\degree}{\mbox{$^{\circ}$}}
\newcommand{\ltish}{\raisebox{-0.4ex}{$\,\stackrel{<}{\scriptstyle\sim}$}}
\newcommand{\vs}{{\em vs\/}}
\newcommand{\bin}[2]{\left(\begin{array}{c} #1 \\ #2\end{array}\right)}
                        
\draft

\title{Coherent dynamics of Bose-Einstein condensates in high-finesse
optical cavities}

\author{Peter Horak${}^1$, Stephen M.\ Barnett${}^1$, and Helmut Ritsch${}^2$}
\address{${}^1$ Department of Physics and Applied Physics,
University of Strathclyde, Glasgow G4 ONG, United Kingdom}
\address{${}^2$ Institut f\"ur Theoretische Physik,
Universit\"at Innsbruck, Technikerstr.\ 25, A-6020 Innsbruck, Austria}
\date{\today}

\maketitle

\begin{abstract}
We study the mutual interaction of a Bose-Einstein condensed gas with a single
mode of a high-finesse optical cavity. We show how the cavity transmission
reflects condensate properties and calculate the self-consistent intra-cavity
light field and condensate evolution. Solving the coupled condensate-cavity
equations we find that while falling through the cavity, the condensate is
adiabatically transfered into the ground state of the periodic optical
potential. This allows time dependent non-destructive measurements on
Bose-Einstein condensates with intriguing prospects for subsequent controlled
manipulation.
\end{abstract}

\pacs{PACS number(s): 3.75.Fi, 3.65.Bz, 42.50.Vk, 42.50.Gy}

\narrowtext



Since the first experimental realisations of Bose-Einstein condensation in
dilute gases \cite{Anderson,Bradley,Davis}, 
the properties and possible applications of condensates in various situations
have been investigated. Most recently, attention has been drawn to the study of
condensates in optical lattices 
\cite{Kozuma,Choi,Jaksch,Berg,Marzlin}
as have been intensely used in the context of laser cooling and trapping of
clouds of noninteracting atoms \cite{Nobel}. But whereas the occupation of the
lattice sites in an optical molasses for a cloud of laser cooled atoms at best
is one atom per ten wells, the atomic densities found in a condensate allow for
multiple occupation of each single well, which gives rise to a variety of new
phenomena.

In this Letter we investigate the case of a condensate falling through a driven
high-finesse optical cavity. The strong coupling of the condensed atoms to the
cavity mode changes the resonance frequency of the cavity which hence is
shifted into or out of resonance with the driving field. Consequently, the
intracavity field intensity is modified and this can easily be measured by
detecting the cavity output field. We show that according to the collective
nature of the condensate this gives a measurable effect even for very low field
intensities and for detunings from the atomic resonance frequency so large that
the spontaneous scattering of photons is negligible. The proposed system should
allow us to perform non-destructive measurements on the condensate. Similar
systems have been used recently to predict amplification of matter waves
\cite{Law} and the appearance of dressed condensates \cite{Goldstein,Moore}.



Let us first introduce in more detail our model system, which is similar to
that used recently to study the effect of a dynamically changing cavity field
on the motion of a single atom \cite{CavPRL,CavPRA,Doherty,Rempe}.

We consider a Bose condensate consisting of $N$ two-level atoms of resonance
frequency $\omega_a$ and spontaneous decay rate $\Gamma$ falling through
an optical cavity. The atom-cavity coupling is
\be
g(x,t)=g_0 \cos(kx) e^{-(v_z t)^2/(2 w^2)},
\ee
for a cavity mode in the form of a standing wave in the longitudinal direction
and a Gaussian with waist $w$ transversally. The condensate is assumed to fall
with constant velocity $v_z$, meaning that we neglect the transverse light
forces on the atoms and the gravitational acceleration in the interaction
region. Two further assumptions have been made at this point. First, the
spatial extension of the condensate has to be small compared to the waist $w$
of the cavity in order to allow a quasi one-dimensional treatment. Second, we
assume that the induced resonance frequency shift of the cavity
\cite{Parkins,Hau} is much smaller  than the longitudinal mode spacing, so that
we can restrict the model to a single longitudinal mode with wavenumber $k$.

The cavity with resonance frequency $\omega_c$ and cavity decay rate $\kappa$
is externally driven by a laser of frequency $\omega$ with pump amplitude
$\eta$ and is treated classically, that is, the intra-cavity field is described
by a (complex) field amplitude $\alpha$. 

As we are interested in the limit where the condensate is
not destroyed by the light field, we will assume a large detuning 
$\Delta_a=\omega-\omega_a \gg \Gamma$ of the driving laser from the atomic
resonance such that the saturation parameter $s=g_0^2/\Delta_a^2\ll 1$.
Moreover, we want the cavity decay to dominate over the spontaneous decay of
{\em all\/} atoms and thus impose the condition 
\be
\kappa \gg N \Gamma s.
\label{eq:c1}
\ee
In this realistically achievable limit we are not only allowed to adiabatically
eliminate the excited state of the atoms but also to completely omit the effect
of atomic decay.

Hence we obtain the equation of motion for the field amplitude
\be
\dot{\alpha}(t) = 
   \left[ i\Delta_c - i N\langle U(\hat{x},t)\rangle -\kappa \right]
   \alpha(t) + \eta
\label{eq:aldot}
\ee
where $\Delta_c=\omega-\omega_c$ is the cavity-pump detuning,  
$U(x,t) = g(x,t)^2/\Delta_a$ the optical potential per photon, and 
``$\langle\dots\rangle$'' denotes the expectation value taken with respect to
the condensate wave function $|\psi(t)\rangle$ at this time. This term
describes the action of the condensate on the cavity field: the refractive
index of the condensate shifts the resonance frequency of the cavity by an
amount of  $N\langle U(\hat{x},t)\rangle$.  (With very high finesse optical
cavities this effect has already been observed even for a single atom 
\cite{Rempe,Mabuchi,Munstermann}, that is, for $N=1$.)  If we require that this
effect significantly changes the intra-cavity field intensity in order to yield
a measurable difference in the cavity output, then the maximum frequency shift
must be of the order of or larger than the cavity line width $\kappa$, which
implies
\be
N g_0^2/\Delta_a \ge \kappa.
\ee
  From this we obtain an order of magnitude for the required detuning $\Delta_a$
which we insert into Eq.~(\ref{eq:c1}) to obtain the following condition for
the cavity parameters:
\be
\frac{Ng_0^2}{\Gamma\kappa} \gg 1.
\ee

The condensate wave function itself obeys a nonlinear Schr\"odinger equation
(known as the Gross-Pitaevskii equation) with the Hamiltonian
\be
H = \frac{\hat{p}^2}{2m} + |\alpha(t)|^2 U(\hat{x},t) +
    N g_{coll} |\psi(\hat{x},t)|^2
\label{eq:ham}
\ee
where the last term describes two-particle collisions between the condensed
atoms and is related to the s-wave scattering length $a$ by 
$g_{coll} = 4\pi\hbar^2 a/m$. This Hamiltonian, together with
Eq.~(\ref{eq:aldot}) for the cavity field, forms a set of coupled nonlinear
equations describing the dynamics of the compound system formed by the
condensate and the optical cavity.

In this Letter we will only consider the special case where the
cavity decay rate $\kappa$ is much larger than the oscillation frequency of
bound atoms in the optical potential of the cavity. In this limit the
intra-cavity field amplitude follows adiabatically the condensate
wavefunction and hence at any time is given by
\be
\alpha(t) = \frac{\eta}{\kappa-i[\Delta_c-N\langle U(\hat{x},t)\rangle]}.
\label{eq:alphaconst}
\ee
Thus, the light intensity of the cavity output, which is proportional to
$|\alpha|^2$, provides information about the condensate wavefunction. In the
following we will investigate this effect in certain special parameter limits.



\begin{figure}[tb]
\vspace{-10mm}
\infig{22em}{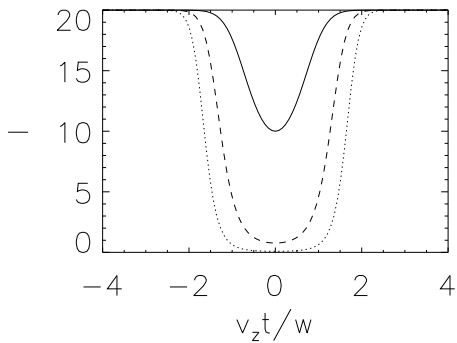}
\vspace{-5mm}
\caption{Cavity photon number $I=|\alpha|^2$ versus time for a condensate
falling through the cavity without being perturbed by
the cavity field (see text for details). The optical potential
depth is given by $NU_0=2\kappa$ (solid line), $10\kappa$ (dashed), and 
$30\kappa$ (dotted), respectively. The pump field is chosen to give rise to
20 photons in the empty cavity and the detuning is $\Delta_c=0$.
}
\label{fig:a}
\end{figure}

Let us first consider the simple case where the cavity field is weak enough and
the interaction time $\tau=w/v_z$ is short enough such that the condensate
wavefunction remains essentially unperturbed (flat on the length scale of an
optical wavelength). In this limit the cavity field can be evaluated
analytically by inserting the frequency shift per atom
\be
\langle U(\hat{x},t)\rangle = \frac{U_0}{2} e^{-(v_z t)^2/w^2} 
\ee
into Eq.~(\ref{eq:alphaconst}), where we have introduced 
$U_0 = g_0^2/\Delta_a$. In Fig.~\ref{fig:a} the
resulting mean cavity photon number $I=|\alpha|^2$ is plotted as
a function of time for different atom numbers in the condensate or,
equivalently, for different optical potentials.

For the parameters chosen in Fig.~\ref{fig:a}, the empty cavity is in resonance
with the driving field but is shifted out of resonance by the presence of the
condensate. Hence the condensate is detected by the {\em absence\/} of light,
which further  reduces the action of the cavity onto the condensate. Therefore
the cavity provides a non-perturbative method of detection. The maximum
resolution of the detection is limited, however, by the cavity waist $w$ and is
not good enough to detect fine structures such as condensate interference
fringes. It might be useful, however, to measure the output of an atom laser as
has recently been realised experimentally \cite{Mewes,Bloch}.  Note also that
this detection scheme only relies on the density of the condensate, not on its
coherence, and thus in principle works as well with an incoherent cloud of
atoms.



Let us now consider the opposite limit of a condensate falling very slowly
through the cavity. We will find that under such conditions the condensate is
adiabatically transfered into the lowest bound state of the optical potential
and hence is strongly localized. In this case we must use the position and time
dependent optical potential
\be
U(\hat{x},t) = U_0 \cos^2(k\hat{x}) e^{-(v_z t)^2/w^2},
\label{eq:U1D}
\ee
so that the condensate at all times feels a periodic optical potential 
with periodicity $\lambda/2$. The condensate wavefunction is conveniently
described in terms of Bloch states
\be
\psi_{n,q}(x) = e^{iqx}\phi_{n,q}(x)
\ee
where the functions $\phi_{n,q}$, $n\geq 0$, are periodic and the Bloch momentum
$\hbar q$ is confined to the interval $[-\hbar k,\hbar k]$. Note that
the coherent interaction of the condensate with the cavity light only couples
Bloch states of different $n$ but leaves states of different $q$ decoupled and
thus $q$ is a conserved quantity.

We are interested in the adiabatic limit corresponding to small transverse
velocities $v_z\ll w/\tau_R$, where $\tau_R$ is the inverse of the recoil
frequency $\omega_R=\hbar k^2/(2m)$. In this limit the coherent time evolution
associated with a potential of the form of Eq.~(\ref{eq:U1D}) maps the Bloch
energy bands $n$ onto the free space momentum intervals 
$[-(n+1)\hbar k, -n\hbar k]$ and $[n\hbar k,(n+1)\hbar k]$. This phenomenon is
known and has been exploited in the context of laser cooling 
\cite{Kastberg,Stecher}. The same effect can be used in our model to transfer a
falling condensate with a transverse momentum distribution confined in
$[-\hbar k,\hbar k]$ into the lowest energy band ($n=0$) of the optical
potential inside the cavity. However, the situation is more complex here than
for the case of laser cooling for two reasons. First, the light intensity
itself depends on the condensate wavefunction and hence on time. Second, the
condensate wavefunction obeys the {\em nonlinear\/} Schr\"odinger equation.
Hence, for any given z-position of the condensate (any given time) the lowest
energy state has to be found by self-consistently solving
Eq.~(\ref{eq:alphaconst}) and the Gross-Pitaevskii equation.

\begin{figure}[tb]
\vspace{-10mm}
\infig{22em}{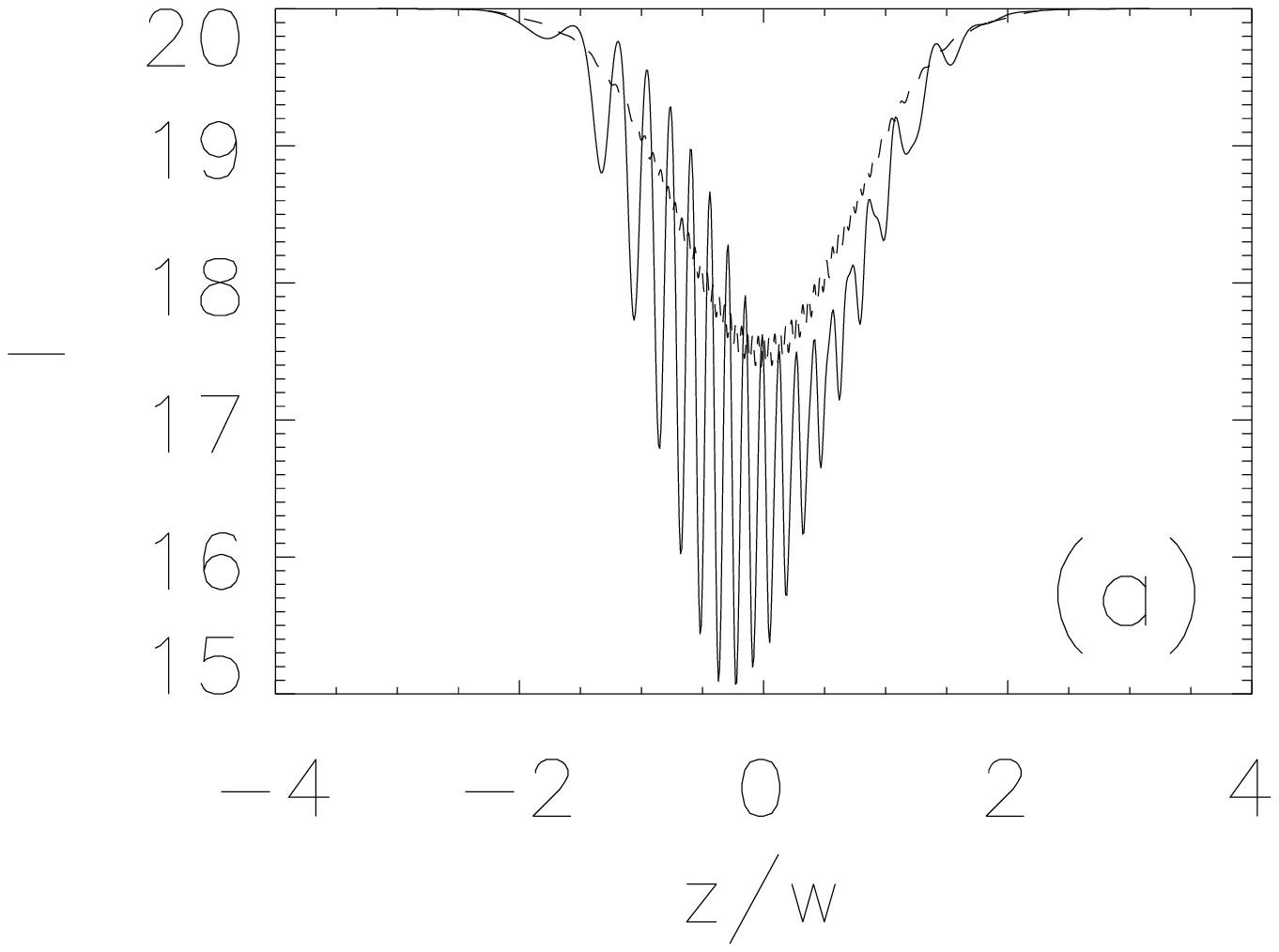}
\vspace{-10mm}
\infig{22em}{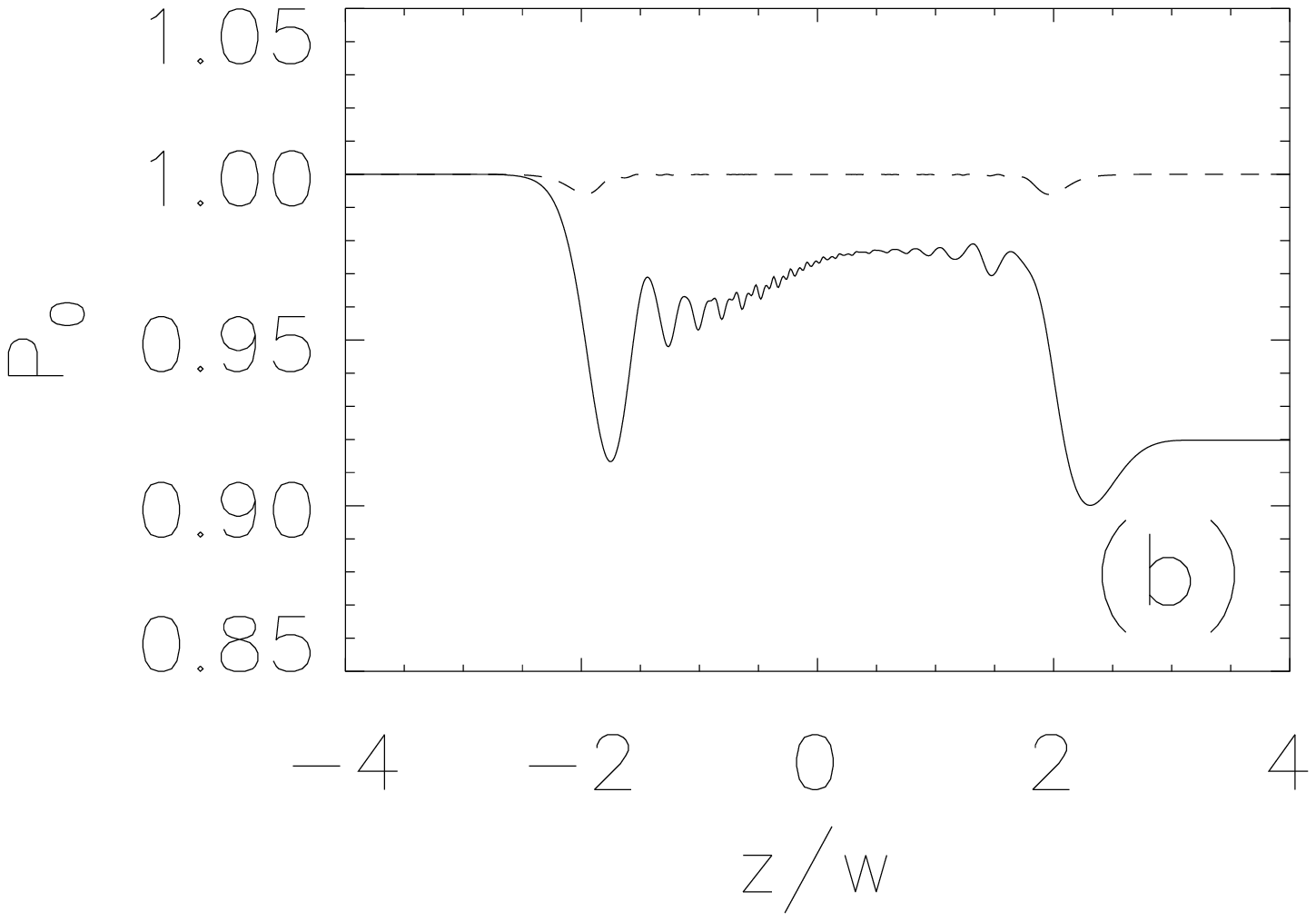}
\vspace{-5mm}
\caption{Adiabatic transfer of the condensate wavefunction to the bound ground
state. $g_{coll}=0$, $\Delta_c=0$, $N U_0=10\kappa$, $v_z=w/(3\tau_R)$ (dashed
curves) respectively $v_z=w/\tau_R$ (solid curves). (a) Intra-cavity photon
number $I=|\alpha|^2$,  (b) population of instantaneous ground state
corresponding to the photon number at any time. The initial condensate
wavefunction is taken as the zero momentum eigenstate  $|p=0\rangle$.
}
\label{fig:b}
\end{figure}

We show in Fig.~\ref{fig:b} the time evolution of the intra-cavity field
intensity $I$ and the overlap $P_0$ of the condensate wavefunction with the
lowest energy band as the condensate is falling through the cavity without
atomic collisions ($g_{coll}=0$). For simplicity we assume that the initial
condensate wavefunction is the free momentum state of zero momentum. The
conservation of the Bloch momentum $q$ then ensures that  at any time only the
Bloch states with $q=0$ are populated and the overlap with the lowest energy
band is given by
\be
P_0 = |\langle\psi|\phi_{n=0,q=0}\rangle|^2.
\ee
For a condensate falling with a velocity $v_z=w/\tau_R$ the transfer of the
free wavefunction into the optical potential is not adiabatic (especially not
on entering and leaving the cavity) and the occupation of the ground state
drops to about 90\%. Hence there is a significant occupation of excited modes
and the corresponding spatial oscillation of the condensate is reflected in the
oscillation of the cavity output. For $v_z=w/(3\tau_R)$, however, nearly all of
the population is transfered into the lowest bound state and accordingly the
oscillations are suppressed.

For the parameters of Fig.~\ref{fig:b}, $U_0>0$ and the condensate is attracted
to the {\em nodes\/} of the light field. Hence, the lowest bound state is
localized at these positions which leads to a much reduced coupling of the
condensate to the cavity and correspondingly to a much smaller frequency shift
of the cavity resonance. This can be easily seen by comparing the results for
the cavity field with the solid line of  Fig.~\ref{fig:a}, which is taken for
the same parameters but a condensate falling so fast that the wavefunction does
not have the time to change due to the presence of the optical potential wells.

\begin{figure}[tb]
\vspace{-10mm}
\infig{22em}{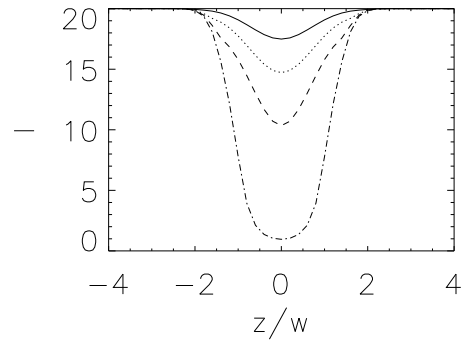}
\vspace{-5mm}
\caption{Photon number corresponding to self-consistent ground state vs
position of condensate in the cavity mode. $Ng_{coll}=0,5,10,20\omega_R$ (from
top to bottom), $\Delta_c=0$,  $N U_0=10\kappa$. (For Rb and Na experiments,
$Ng_{coll}=\omega_R$ corresponds to condensate densities of the order of 30
atoms per potential well, i.e., per $(\lambda/2)^3$.)
}
\label{fig:c}
\end{figure}

Figure~\ref{fig:c} shows the effects of collisions on the intracavity photon
number in the adiabatic regime ($v_z \ll w/\tau_R$). The curves are derived
from the self-consistent solution of Eq.~(\ref{eq:alphaconst}) and the
Gross-Pitaevskii equation for different values of $g_{coll}$. For increasing
$g_{coll}$ the atoms in the condensate increasingly repell each other, 
counteracting the confining effect of the optical potential. The wavefunction
becomes broader and the coupling of the condensate to the cavity stronger,
which in turn leads to larger shifts in the cavity resonance frequency and a
reduced cavity field intensity. Hence, the decrease in the cavity output
provides a direct measure of the atom-atom interaction within the condensate.
Note that this could be used for {\em in situ\/} measurements of Feshbach
resonances, if one manipulates the s-wave scattering length by applying a
magnetic field \cite{Inouye,Courteille,Tiesinga,Kagan}.



\begin{figure}[tb]
\vspace{-10mm}
\infig{22em}{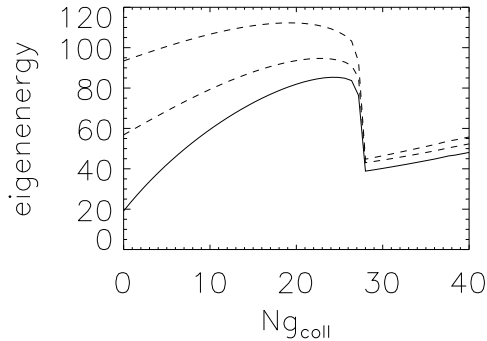}
\vspace{-5mm}
\caption{Eigenenergy of the self-consistent ground state
(solid curve) and of the first two condensate excitation states (dashed curves)
versus collision parameter $g_{coll}$ for
$\Delta_c=0$, $NU_0=10\kappa$.
Eigenenergies and $g_{coll}$ are given in units of $\omega_R$.
}
\label{fig:d}
\end{figure}

Finally, we plot in Fig.~\ref{fig:d} the energy of the self-consistent ground
state and of the two lowest excitations. The latter are calculated in lowest
order perturbation theory, that is, using the self-consistent ground state for
the collisional term of the Hamiltonian (\ref{eq:ham}). The energy difference
between these states is the one which is responsible for the oscillation of the
cavity output in the case of non-perfect adiabatic transfer of the condensate
wavefunction into the lowest bound state as shown previously in
Fig.~\ref{fig:b}.

For $g_{coll}=0$ the ground state is strongly localized at the antinodes of the
cavity leading to a large intracavity field intensity. Hence the lowest three
eigenstates are well approximated by harmonic oscillator states and the
frequency difference between the ground state and the first excited state is
the same as the frequency difference between the first and the second excited
state. For increasing $g_{coll}$ the internal (collisonal) energy of the
condensate increases. Simultaneously the effective potential (optical potential
plus collisional term) for the excited modes changes its shape from a harmonic
oscillator to a potential well with a nearly flat bottom. This is due to the
well-known fact that the ground state of the Gross-Pitaevskii equation becomes
proportional to the negative confining potential if the collisional term
dominates over the kinetic energy term in the Hamiltonian (Thomas-Fermi limit).
Hence the excited modes are those of a potential which looks more like a box in
this limit and the level spacing increases between more highly excited modes.
However, as the wavefunction broadens with increasing $g_{coll}$ the cavity
field intensity decreases and above a certain value  ($Ng_{coll}\approx
28\omega_R$ for the parameters in Fig.~\ref{fig:d}) no bound state exists in
the optical potential. Hence the eigenfunctions above this critical value
resemble free space momentum states which leads to the significant change in
the behaviour of the spectrum of excitation energies.



In conclusion we have shown that a high-quality optical cavity provides a
powerful tool to investigate properties of Bose-Einstein condensates in a
non-destructive way. For slow initial velocities the condensate is
adiabatically transfered into the self-consistent ground state of the optical
potential, which contains ample information on condensate properties.

In addition, the system suggests many interesting applications. Using
cavities with a decay rate $\kappa$ of the order of the condensate vibrational
frequencies in the potential, the finite response time of the cavity implies a
damping (or amplification) of the condensate oscillations as it has been shown
for cooling and trapping of a single atom \cite{CavPRL,CavPRA,Rempe}. By
changing the intensity and/or the frequency of the driving laser depending on
the cavity output one gets a handle for the controlled non-destructive {\em in
situ\/} manipulation of the condensate wavefunction. Similarly, changing the
magnitic field during the condensate passage should allow a direct measurement
of the effective scattering length and the relaxation dynamics.

This work was supported by the United Kingdom EPSRC, the Austrian Science
Foundation FWF (P13435), and the European Comission TMR network
(FRMX-CT96-0077).



\begin{thebibliography}{99}

\vspace{-10mm}
\bibitem{Anderson} M.\ H.\ Anderson {\em et al.}, Science {\bf 269}, 198 (1995).

\bibitem{Bradley} C.\ C.\ Bradley {\em et al.}, Phys.\ Rev.\ Lett.\ {\bf 75}, 
   1687 (1995).

\bibitem{Davis} K.\ B.\ Davies {\em et al.}, Phys.\ Rev.\ Lett.\ {\bf 75}, 
   3969 (1995).

\bibitem{Kozuma} M.\ Kozuma {\em et al.}, Phys.\ Rev.\ Lett.\ {\bf 82}, 
   871 (1999).

\bibitem{Choi} D.-I.\ Choi and Q.\ Niu, Phys.\ Rev.\ Lett.\ {\bf 82}, 
   2022 (1999).

\bibitem{Jaksch} D.\ Jaksch {\em et al.}, Phys.\ Rev.\ Lett.\ {\bf 81}, 
   3108 (1998).

\bibitem{Berg} K.\ Berg-S{\o}rensen and K.\ M{\o}lmer, Phys.\ Rev.\ A 
   {\bf 58}, 1480 (1998).

\bibitem{Marzlin} K.-P.\ Marzlin and W.\ Zhang, Phys.\ Rev.\ A {\bf 59}, 
   2982 (1999).

\bibitem{Nobel} S.\ Chu, C.\ Cohen-Tannoudji, W.\ D.\ Phillips, Nobel lectures,
   Rev.\ Mod.\ Phys.\ {\bf 70}, 685-742 (1998).

\bibitem{Law} C.\ K.\ Law and N.\ P.\ Bigelow, Phys.~Rev.~A {\bf 58}, 4791
   (1998).

\bibitem{Goldstein} E.\ V.\ Goldstein, E.\ M.\ Wright, and P.\ Meystre, 
   Phys.~Rev.~A {\bf 57}, 1223 (1998).

\bibitem{Moore} M.\ G.\ Moore and P.\ Meystre, Phys.~Rev.~A {\bf 59}, R1754
   (1999); M.\ G.\ Moore, O.\ Zobay, and P.\ Meystre, e-print cond-mat/9902293.

\bibitem{CavPRL} P.\ Horak {\em et al.}, Phys.~Rev.~Lett.\ {\bf 79}, 4974
   (1997).

\bibitem{CavPRA} G.\ Hechenblaikner {\em et al.}, 
   Phys.~Rev.~A {\bf 58}, 3030 (1998).

\bibitem{Doherty} A.\ C.\ Doherty {\em et al.}, Phys.~Rev.~A {\bf 56}, 833 
   (1997).

\bibitem{Rempe} P.\ M\"unstermann {\em et al.}, Phys.~Rev.~Lett.\ {\bf 82}, 
   3791 (1999).

\bibitem{Parkins} A.\ S.\ Parkins and D.\ F.\ Walls, Phys.~Rep.\ {\bf 303}, 
   1 (1998).

\bibitem{Hau} L.\ V.\ Hau {\em et al.}, Nature {\bf 397}, 594 (1999).

\bibitem{Mabuchi} H.\ Mabuchi {\em et al.}, Opt.\ Lett.\ {\bf 21}, 1393 (1996);
   C.\ J.\ Hood {\em et al.}, Phys.\ Rev.\ Lett.\ {\bf 80}, 4157 (1998).

\bibitem{Munstermann} P.\ M\"unstermann {\em et al.}, Opt.\ Commun.\ 
   {\bf 159}, 63 (1999).

\bibitem{Mewes} M.-O.\ Mewes {\em et al.}, Phys.\ Rev.\ Lett.\ {\bf 78}, 
   582 (1997).

\bibitem{Bloch} I.\ Bloch, T.\ W.\ H\"ansch, and T.\ Esslinger, 
   Phys.\ Rev.\ Lett.\ {\bf 82}, 3008 (1999).

\bibitem{Kastberg} A.\ Kastberg {\em et al.}, Phys.\ Rev.\ Lett.\ {\bf 74}, 
   1542 (1995).

\bibitem{Stecher} H.\ Stecher {\em et al.}, Phys.\ Rev.\ A {\bf 55}, 
   545 (1997).

\bibitem{Inouye} S.\ Inouye {\em et al.}, Nature {\bf 392}, 151 (1998).

\bibitem{Courteille} P.\ Courteille {\em et al.}, Phys.\ Rev.\ Lett.\ 
   {\bf 81}, 69 (1998).

\bibitem{Tiesinga} E.\ Tiesinga, B.\ J.\ Verhaar, and H.\ T.\ C.\ Stoof, 
   Phys.\ Rev.\ A {\bf 47}, 4114 (1993).

\bibitem{Kagan} Y.\ Kagan, E.\ L.\ Surkov, and G.\ V.\ Shlyapnikov, 
   Phys.\ Rev.\ Lett.\ {\bf 79}, 2604 (1997).


\end{thebibliography}
\end{document}